\documentclass[a4,paper,12pt]{article}
\usepackage[english]{babel}
\usepackage[utf8]{inputenc}
\usepackage{amsmath, enumerate,anysize,upgreek,fixmath}
\usepackage{graphics}
\usepackage{graphicx}  
\usepackage{color}
\usepackage{caption}
\usepackage{longtable}
\usepackage{cancel}
\usepackage{stackrel}

\usepackage{amssymb,latexsym}

\usepackage{amsfonts,amscd,fancybox,array}
\usepackage{here} %

\begin{document}

\vspace{0.5cm}
\begin{center}
\Large\textbf{Semiparametric Modelling of Cancer Mortality Trends in Colombia}\\
\vspace{0.5cm}
\small{\textsc{Lina Angélica Buitrago Reyes\footnote{Universidad Nacional de Colombia. E-mail: labuitragor@unal.edu.co.}, Juan Sosa\footnote{Universidad Nacional de Colombia. E-mail: jcsosam@unal.edu.co.}, Cristian Andrés González Prieto\footnote{Instituto Nacional de Cancerología. E-mail: cragonzalezpr@unal.edu.co.}}}
\end{center}
\vspace{0.5cm}

\begin{abstract}
In this paper, we compare semiparametric and parametric model adjustments for cancer mortality in breast and cervical cancer and prostate and lung cancer in men, according to age and period of death.
Semiparametric models were adjusted for the number of deaths from the two localizations of greatest mortality by sex: breast and cervix in women; prostate and lungs in men.  Adjustments in different semiparametric models were compared; which included making adjustments with different distributions and variable combinations in the parametric and non-parametric part, for localization as well as for scale.  Finally, the semiparametric model with best adjustment was selected and compared to traditional model; that is, to the generalized lineal model with Poisson response and logarithmic link.
Best results for the four kinds of cancer were obtained for the selected semiparametric model by comparing it to the traditional Poisson model based upon AIC, envelope correlation between estimated logarithm rate and real rate logarithm.  In general, we observe that in estimation, rate increases with age; however, with respect to period, breast cancer and stomach cancer in men show a tendency to rise over time; on the other hand, for cervical cancer, it remains virtually constant, but for lung cancer in men, as of 2007, it tends to decrease. 
\end{abstract}

\section{Introduction}
Cancer, which can occur at any age, is a disease characterized by rapid, abnormal-cell multiplication in any bodily organ \cite{OMS}.  Carcinogenic cells invade adjacent organs and are later propagated in other bodily sites (metastasis), thereby altering vital functions, and in most cases, causing death.  Major cancer risk factors include tobacco, poor diet \cite{NCI} and chronic infections (bacteria, virus or parasites) \cite{Ohshima}.

Furthermore, cancer mortality is ranked among the highest among all diseases world-wide. In the year 2015, according to the World Health Organization (WHO), 8.8 million cancer-related deaths were registered in the world \cite{OMS}; among the most common were lung, hepatic, gastric, colorectal and breast cancers.  National health care institutes around the world monitor cancer mortality as part of their follow up efforts on local cancer control policies and on cancer-mitigation strategies: these include improving access to early detection programs and improving diagnosis and treatment; as well as identifying the most frequent cancers, thereby allowing for strategies to be developed that will have an on specific cancers. 

Between the years 2000 and 2006, in Colombia, 203907 cancer deaths were registered; the most frequent cancers were: stomach, lung, cervical, colon and breast \cite{Atlas}. 

Worldwide, different methods have been utilized to model temporal cancer mortality tendencies.  For example, in 2009, Cabanes et al., modeled mortality for three types of cancer: breast, ovarian and cervical among the female population in Spain from 1980 to 2006.  They used the generalized lineal model with Poisson response by taking age into account to evaluate change in general and specific mortality rates through time \cite{Cav}.  However, nothing has been reported on the virtue of said adjustment, nor on the evaluation of the assumptions.

In 2010, Clèries et al., used the Bayesian age-drift model with Poisson response and different priors according to adjusted model in order to model testicle cancer tendency in men aged 15 to 74 years in Spain from 2005 to 2019 \cite{Cle}.  Additionally, they adjusted an autoregressive APC model to estimate projections in the years for which they had no information and selected the best model for each case based upon deviance information criterion (DIC).

For prostate cancer mortality data in Norway from 1980 to 2007, Kv\aa le et al. adjusted a joinpoint regression model to identify lineal changes in mortality tendency for this type of cancer; however, no results are shown for model's quality \cite{kv}.

In 2012, Guo \& Li modeled the esophagus cancer mortality tendency in China between 1987 and 2009 \cite{Guo}.  To be able to do so, they modeled age, period and birth cohort effect with a generalized additive model (GAM) with Poisson response, in which non-lineal association between predictors is modeled by means of non-parametric techniques.  Model adjustment for this case was good; however, assumption evaluations for this family of models was not reported.  

In Colombia, cancer mortality rates in the Cancer Mortality Atlas \cite{Atlas} were modeled with generalized lineal models, wherein the number of deaths is assumed to be derived from a Poisson distribution whose rate may be explained by age and period, which in this case are drawn from the 2000 to 2006 period.  The Colombian National Statistics Agency's (DANE) official registries provide the sources for this death-related data.  However, these models are not the most adequate for modelling this type of data since the may incur in overdispersion \cite{Berk}; hence, it is necessary to search for a new methodology capable of overcoming the problems at hand and of improving model estimations and general adjustment. 

In this article, we propose the use of semiparametric models \cite{Van} for cancer mortality data modelling, and we compare these to the Poisson models that are in general use. This comparison is made for assumption evaluations, Akaike criteria and for the relation between rate estimation and real rate for 4 types of cancer: the two most frequent (stomach and lung) in men and the two most frequent (breast and cervical) in women.  These models allow for semiparametric modelling of data median and bias. 

The methods section includes descriptions of data used to compare models, as well as a brief description on the semiparametric models used. The results section shows data adjustments via semiparametric models and general lineal Poisson model; and in the discussion section, we address relevance of use of semiparametric models and the diverse advantages they hold over Poisson models.

\section{Materials and Methods}
Mortality data for the 1994-2013 period were obtained from Colombian National Statistics Agency (DANE) official registries; cause of death is coded in accordance with the International Classification of Diseases (ICD), from 1994 to 1997 as ICD-9, and later as ICD-10 \cite{WHO}.  With regards to population, DANE projections according to age and sex were used \cite{DANE}.

The two cancers in men with highest mortality rates, prostate and lung, and in women, breast and cervical, were selected for this study.  The semiparametric models modelled median and bias data for each cancer studied \cite{Van}. Death counts were made by age and year of death (period); therefore, explanatory variables were age group and period midpoint.  Simultaneously, by using the same explanatory variables, generalized lineal models with Poisson response were adjusted and models were compared on the basis of adjustment measurements such as Akaike information criterion and correlation between estimated rate logarithm and observed rate logarithm.

\subsection{Semiparametric models for modelling median and skewness}

Let $T_1,T_2,...,T_n$ assumend to be independent random variables, general model structure is \cite{Van}: 

\begin{equation}
    T_k=\eta_k\xi_k^{\sqrt{\phi_k}}\text{   ,     } k=1,...,n  \label{mod1}
\end{equation}

where $\eta_k$ and $\phi_k>0$ are the median and skewness of $T_k$, $\xi_1,...,\xi_n$ random, independent multiplicative errors, with log-symmetric distribution:

$$\xi_k\stackbin{i.i.d.}\sim\mathcal{LS}(1,1,g(.))$$

Taking logarithm in (\ref{mod1}):

\begin{align}
    \log T_k&=\log\eta_k+\sqrt{\phi_k}\log\xi_k \\ 
    Y_k&=\mu_k +\sqrt{\phi_k}\epsilon_k \text{   ,     } k=1,...,n \label{mod2}
\end{align}

With $Y_k=\log T_k$, $\mu_k=\log\eta_k$ is the $Y_k$ localization parameter, $\phi_k$ is the $Y_k$ dispersion parameter and $\epsilon_k=\log\xi_k$ with symmetric distribution around zero and dispersion 1: $(\epsilon_k\stackbin{i.i.d.}\sim\mathcal{S}(0,1,g(.)))$. 

It is assumed that $\eta_k$ and $\phi_k$ follow the form: 
$$\eta_k=\eta(\boldsymbol{x}_k,\boldsymbol{\beta})$$
$$\log\phi_k=\boldsymbol{w}_k^t\boldsymbol{\gamma}+f(\boldsymbol{b}_k)$$

where $\boldsymbol{x}_k$,$\boldsymbol{w}_k$ y $\boldsymbol{b}_k$ are explanatory variables, $\boldsymbol{\beta}$ and $\boldsymbol{\gamma}$ are parameter vectors, $\log(\eta(\boldsymbol{x}_k,\boldsymbol{\beta}))$ are $f(\boldsymbol{b})$ continue and twice differentiable non-parametric function $\boldsymbol{\beta}$ and $\boldsymbol{b}$ respectively. In this specific case, $T$ is the number of deaths, the explanatory variables vectors contain age group and death period midpoint, and as offset, the population logarithm. 

\subsection{Generalized lineal model with Poisson response}

This type of model is the most commonly used in modelling number of cases in a particular event.  In this case, the adjusted models for each type of cancer employ the form:

\begin{equation}
    \log\mu_i=\log p_i+\boldsymbol{x}_i^t\boldsymbol{\beta}    \label{mod}
\end{equation}

where $\mu_i=E(T_i)$, $T_i$ as the number of deaths, $\boldsymbol{x}_i$ the explanatory variables vector, which has the midpoint of age group and of death period, and $p_i$ the exposed age group population and corresponding period. 

All calculations were made with \verb"R" software \cite{R}; semiparametric models were specifically adjusted with the \verb"ssym" packet \cite{Van_R}.

\section{Results:}
\subsection{Breast cancer}
For breast cancer mortality, the semiparametric model with best adjustment was the one adjusted with contaminated normal distribution whose localization parameter modelling was based upon natural cubic splines for age as well as for period, and scale parameter was based upon natural cubic splines for age.  When compared to traditional model, the semiparametric model showed better adjustment, since much lower AIC was obtained; good adjustment was observed in envelopes, as well as in median and bias (Figure \ref{Fig1}), and correlation between estimations and real data was greater for same (Table \ref{tab:mama}, Figure \ref{Fig2}).

In analyzing semiparametric model results, on the basis of median adjustment, it was estimated that breast cancer mortality goes up rapidly between ages 20 to 50 years; later, it goes down slightly, until age 75, when it then begins to go back up.  Regarding period, it was observed that this increased during the entire period under study; however, said increase is greater as of 2003.  Bias estimation goes down until 55 years of age, when it then begins to slowly go back up (Figure \ref{Fig3}). 

\subsection{Cervical cancer}

For cervical cancer mortality, the semiparametric model with best adjustment was the normal one. Localization adjustment was based upon natural cubic splines for age as well as period, and scale parameter was based upon natural cubic splines for age.  When compared to traditional Poisson model, the semiparametric model showed better adjustment, since much lower AIC was obtained; additionally, good adjustment in both median and bias envelopes was observed (Figure \ref{Fig2}); whereas, for the Poisson model, residuals went completely outside confidence bands, and correlation between estimations and real data was greater for same (Table \ref{tab:cuello}, Figure \ref{Fig2}).

In analyzing semiparametric model results based upon median adjustment, age behavior is similar to that for breast cancer mortality, since rapid increase is observed; this time, from ages 15 to 45 years; later it goes down very slightly, until age 75 years, when it begins to go back up.  Regarding period, parabolic- shaped mortality estimation is almost constant, reaching its peak in 2004.  With regards to bias estimation, it goes down until age 55, when it then begins to slowly go back up (Figure \ref{Fig3}).

\subsection{Gastric cancer}
For gastric cancer mortality, the semiparametric model with best data adjustment was the normal one.  Localization parameter adjustment was based upon natural cubic splines for both age and period. Scale parameter was adjusted with natural cubic splines.  When semiparametric model was compared to traditional Poisson model, the former showed better data adjustment, which can be seen in AIC comparisons for each one: the semiparametric model has a much smaller AIC.  Additionally, median and bias envelopes suggest that adjustment is better than it would be with the Poisson model, in whose envelope residuals move outside the bands (Figure \ref{Fig1}).  Correlation between model's estimated and real values is greater in the semiparametric model than in the Poisson model (Table  \ref{tab:esth}). 

Based upon adjustment the semiparametric model makes for median, behavior similar to that for mortality in previous cancer types can be observed: i.e., accelerated increase as age goes up. Furthermore, it is apparent that mortality increases as period expands. For bias, maximum value is found at age 20, after which time it then goes down.  (Figure \ref{Fig3}).

\subsection{Lung cancer}
For lung cancer mortality in men, the semiparametric model with best data adjustment is the one that uses exponential power distribution.  P-splines were used for scale parameter and for age and period localization parameter modelling.

\begin{figure}
    \centering
    \includegraphics[width=130mm]{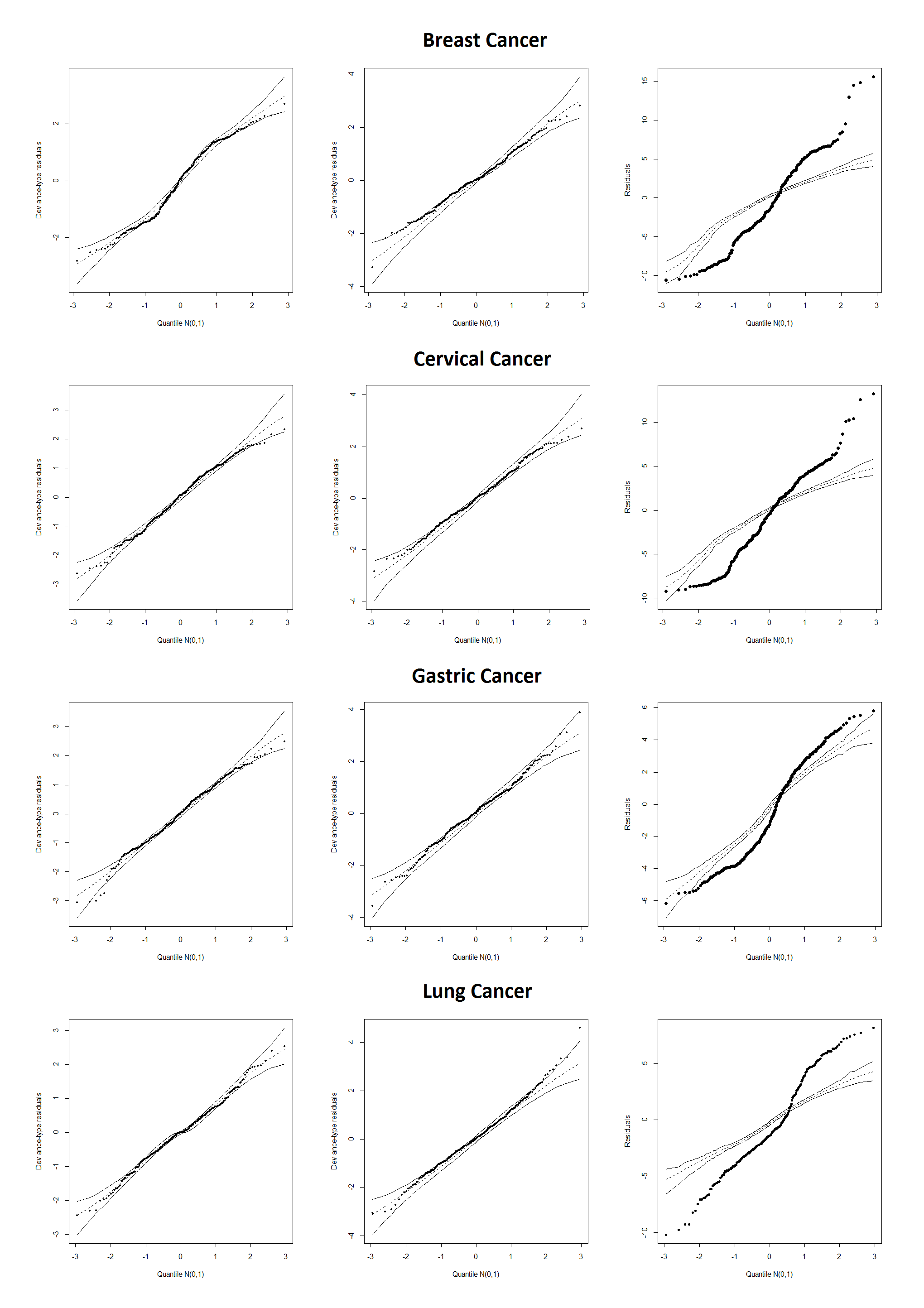}
    \caption{Left: Envelope for localization parameter model in semiparametric model.  Center: Envelope for bias parameter in semiparametric model.  Right: Envelope for Poisson model.}
    \label{Fig1}
\end{figure}

\begin{figure}
    \centering
    \includegraphics[width=130mm]{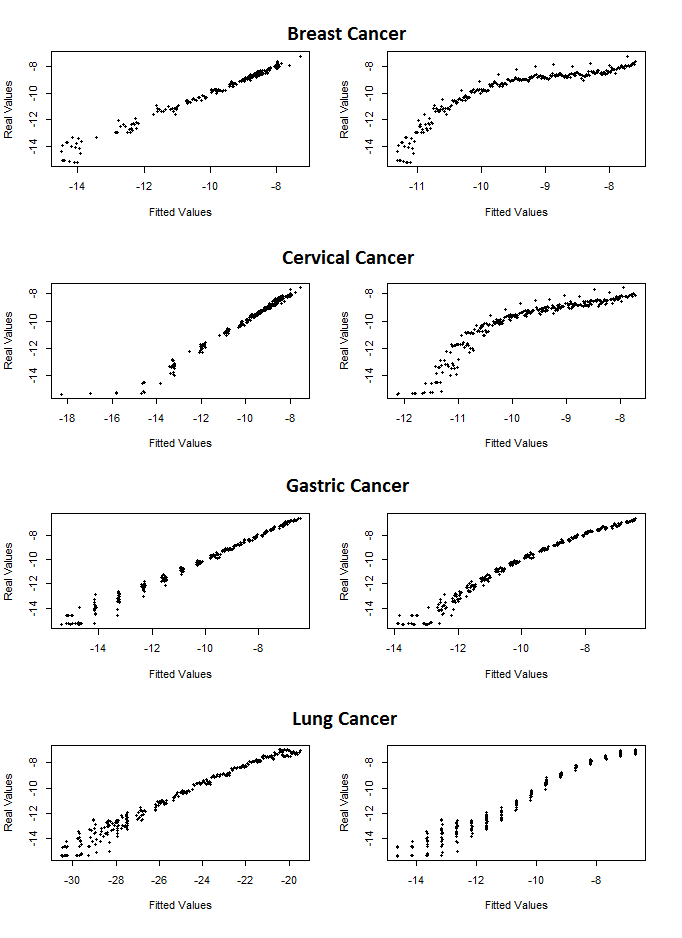}
    \caption{Estimated rate vs. Observed rate.  Left: Semiparametric model.  Right: Poisson model.}
    \label{Fig2}
\end{figure}

When comparison was made for mortality modelling between the semiparametric model and the Poisson model, it was apparent that the former had better data adjustment than the latter.  Semiparametric model AIC is less than that in the Poisson model (Table \ref{tab:pulmon}).  When examining envelope graphs, the resultant Poisson model residuals move outside the bands; however, the same does not happen with semiparametric model residuals (Figure \ref{Fig1}).  Correlation between estimated values and real values is higher in the semiparametric model than in the Poisson model, as Figure \ref{Fig2} illustrates. 

\begin{figure}
    \centering
    \includegraphics[width=130mm]{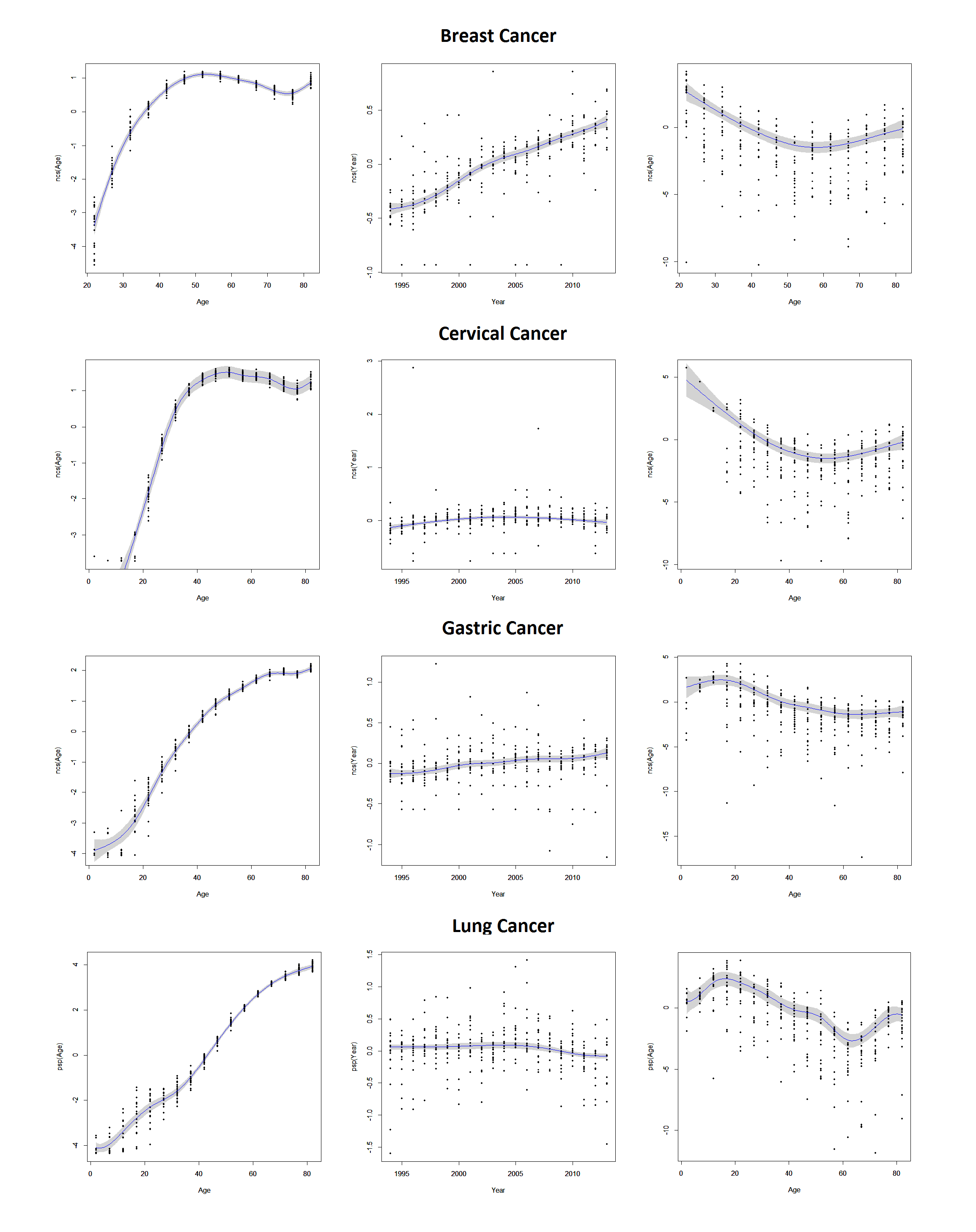}
    \caption{Graphs of non-parametric model components.  For submodel localization parameter, those corresponding to age and period located in left and central columns, respectively.  In right column, submodel skeness age-curve.}
    \label{Fig3}
\end{figure}

Semiparametric model median adjustment reveals the fact that mortality goes up as age advances; however, this remains constant and tends to go down as years go by.  For bias, maximum value is at 20 years of age and minimum value at 60 years of age (Figure \ref{Fig3}).

\begin{table}[H]
    \centering
\begin{tabular}{p{2cm}p{2cm}ll|llll}\hline
\multicolumn{4}{c}{\textbf{Semiparametric Model}}& \multicolumn{4}{c}{\textbf{Poisson Model}} \\\hline
    \textbf{Median}  &  &  &  &  &  &      \\\cline{1-4}

Parametric &  Estimate & Std.Err & $p-$value &  &  Estimate & Std.Err & $p-$value \\
 \textit{Intercept} & 4,3114 & 0,0156 & $<0,005$ &  Intercept & -40,79 & 1,845 & $<0,005$ \\
           &  &  &  &  age &  0,058 &  0,000308 &  $<0,005$ \\
           &  &  &  &  period &  0,014  &  0,0009205  &  $<0,005$ \\
Non parametric & Smooth param &  d.f  &  $p-$value  &  &  &  &  \\
  \textit{nsc(age)} &  706,7  &  7,846  &  $<0,005$  &  &  &  &  \\
\textit{ncs(period)} &  2376,9  &  5,607  &  $<0,005$  &  &  &  & \\\cline{1-4}
  \textbf{Skewness} &  &  &  &  &  &  & \\\cline{1-4}
Parametric &  Estimate  &  Std.Err  &  $p-$value &  &  &  &  \\
 \textit{Intercept} &    -5,9546 &  0,1004  &  $<0,005$ &  &  &  & \\
Non parametric &  &  &  &  &  &  & \\
  \textit{nsc(age)} &  3547  &  3,23  &  $<0,005$ &  &  &  &  \\\hline
       AIC &  -263,33  &  &  &  AIC  &  8832,1  &  &  \\
      $\rho$ &  0,9919 &  &  &  $\rho$ &  0,8831 &  &  \\\hline
\end{tabular}  
    \caption{Mortality models for breast cancer.}
    \label{tab:mama}
\end{table}

\begin{table}[H]
    \centering
    \begin{tabular}{p{2cm}p{2cm}ll|llll}\hline

\multicolumn{4}{c}{\textbf{Semiparametric Model}}& \multicolumn{4}{c}{\textbf{Poisson Model}} \\\hline
    \textbf{Median} &  &  &  &  &  &  &  \\ \cline{1-4}
Parametric & Estimate & Std.Err & $p-$value &  & Estimate & Std.Err & $p-$value \\
 \textit{Intercept} & 3,6658 & 0,0548 &$<0,005$ & \textit{Intercept} & 32,1867 & 1,9513 &$<0,005$ \\
           &  &  &  & age & 0,0548 & 0,0003 & $<0,005$ \\
           &  &  &  & period & -0,0223 & 0,00097 & $<0,005$ \\
Non parametric & Smooth param &  d.f & $p-$value &  &  &  &  \\
\textit{nsc(age)} & 2261 & 7,65 & $<0,005$ &  &  &  & \\
\textit{ncs(period)} & 35689 & 2,937 & $<0,005$ &  &  &   &  \\\cline{1-4}
  \textbf{Skewness} &            &            &            &            &            &            &            \\\cline{1-4}
Parametric &  Estimate & Std.Err & $p-$value &            &            &            &            \\
 Intercept & -3,5951 &   0,0973 & $<0,005$ &            &            &            &            \\
Non parametric &    &   &   &   &            &            &            \\
  \textit{nsc(age)} &   7538 &  3,252 & $<0,005$ &            &            &            &            \\\hline
       AIC &   -305,784 &            &            &        AIC &     7416,2 &            &            \\
      $\rho$ &     0,9898 &            &            &       $\rho$ &     0,9077 &            &            \\\hline
\end{tabular}  
    \caption{Mortality models for cervical cancer.}
    \label{tab:cuello}
\end{table}

\begin{table}[H]
    \centering
    \begin{tabular}{p{2cm}p{2cm}ll|llll}\hline

\multicolumn{4}{c}{\textbf{Semiparametric Model}} & \multicolumn{4}{c}{\textbf{Poisson Model}} \\\hline
    \textbf{Median} &  &  &  &  &  &  & \\\cline{1-4}
Parametric &  Estimate & Std.Err & $p-$value &  & Estimate & Std.Err & $p-$value \\
 \textit{Intercept} & 4,0002 & 0,0179 &$<0,005$ &  \textit{Intercept} & 17,4025113 &  1,5337669 & $<0,005$ \\
           &   &  &   & \textit{age} &  0,0912659 &  0,0002885 & $<0,005$ \\
           &   &  &   & \textit{period} & -0,0157019 &  0,0007656 & $<0,005$ \\
Non parametric & Smooth param &  d.f & $p-$value &  &  &  &  \\
  \textit{nsc(age)} & 718,7 & 7,91 & $<0,005$ &  &  &  &  \\
ncs(period) &  2285,6 & 5,422 & $<0,005$ &  &  &  &  \\\cline{1-4}
  \textbf{Skewness} &   &  &   &   &   &   &    \\\cline{1-4}
Parametric &   Estimate & Std.Err & $p-$value &  &  &  &  \\
 \textit{Intercept} & -3,7278 & 0,0899 & $<0,005$ &  &  &  & \\
Non parametric & Smooth param &  d.f & $p-$value &  &  &  &  \\
  \textit{nsc(age)} &  809,4 & 5,627 & $<0,005$ &  &  &  &  \\\hline
       AIC &   -237,782 &  &  & AIC & 4448,3 &  &  \\
      $\rho$ & 0,9951 &  &  & $\rho$ & 0,9841 & &   \\\hline
\end{tabular}  
    \caption{Mortality models for gastric cancer in men.}
    \label{tab:esth}
\end{table}

\begin{table}[H]
    \centering
    \begin{tabular}{p{2cm}p{2cm}ll|llll}\hline

\multicolumn{4}{c}{\textbf{Semiparametric Model}} & \multicolumn{4}{c}{\textbf{Poisson Model}}  \\\hline
    \textbf{Median} &  &  &  &  &  &  &  \\\cline{1-4}
Parametric &  Estimate &  Std.Err & $p-$value &   & Estimate &  Std.Err & $p-$value \\
 \textit{Intercept} & -11,089 &  0,0218 &  $<0,005$ & \textit{Intercept} & 11,99 &  1,747 & $<0,005$ \\
     &   &   &   &  \textit{age} &  0,099 &  0,3442 & $<0,005$ \\
     &   &   &   &  \textit{period} & -0,0014 & 0,8719 &  0,103 \\
Non parametric & Smooth param &  d.f & $p-$value &  &  &  &   \\
 \textit{psp(age)} & 586,851 & 3,121 & $<0,005$ &   &   &   &   \\
\textit{psp(period)} & 1,111 & 8,999 & $<0,005$ &   &   &   &   \\\cline{1-4}
  \textbf{Skewness} &  &  &  &   &   &   &   \\\cline{1-4}
Parametric & Estimate &  Std.Err & $p-$value &   &   &   &  \\
\textit{Intercept} &  71,237 & 2,698 & 0,008 &  &   &   &  \\
Non parametric & Smooth param & d.f & $p-$value  &   &   &   &   \\
  \textit{psp(age)} &  0,00272 & 7,578 &  $<0,005$ &    &   &   & \\
\textit{psp(period)} & 0,05815 & 8,793 &  $<0,005$ &    &   &   & \\\hline
       AIC & -67,937 &  &  & AIC & 6199,9 &  &  \\
      $\rho$ &  0,9881 &  &  & $\rho$ & 0,9844 &  &  \\\hline
    \end{tabular}
    \caption{Mortality models for lung cancer in men.}
    \label{tab:pulmon}
\end{table}

\section{Discussion}
Cancer-mortality tendency modelling is relevant in analyzing the impact that cancer control strategies may have. As these strategies are worked out, tendencies in cancer are revealed, and modelling can play an essential role in the ongoing implementation of said strategies.

Mortality rates, as may be expected, increase with age. However, in the cases of breast cancer and cervical cancer, rates go up until approximately age 50, when mortality rate then begins to slightly go down; a fact that can be explained by reduced estrogen production during this phase women's lives. (Figure \ref{Fig3}) \cite{Pike}\cite{Marchant}.

Mortality tendency for breast cancer shows no apparent decrease over time; on the contrary, in spite of multiple early detection campaigns that have been carried out \cite{INC}, it tends to rise (Figure \ref{Fig3}). For cervical cancer, the tendency remains constant. For stomach cancer in men, mortality tendency goes up slightly over time; in contrast, lung cancer in men slightly decreases during same period.  

Regarding model comparisons, it was shown that traditional model adjustments were of lesser quality, due to the fact that greater AIC were obtained, and that problems with residuals' overdispersion were encountered in said models (Figure \ref{Fig1}); as Figure \ref{Fig3} corroborates.  Another finding that verifies the foregoing was the higher linearity and lower variance between prediction and real data (See Figure \ref{Fig2}).  Taking the above into account, it is apparent that the relation between age, period and cancer mortality rate, in this case, is not lineal; therefore, it is necessary to consider methodologies that include at least some non-parametric part that will allow for non-linearity to be accounted for.  

On the other hand, non-parametric model adjustments suffer from the disadvantage of not allowing for parameter estimation, nor for summary measures such as annual average percentage change; on the contrary, the entire description is graphic.

The semiparametric model methodology used in this study allows for the utilization of different types of positive bias distribution in making adjustments.  For breast cancer, the contaminated normal type was used; for cervical and stomach cancers, the normal type; and for lung cancer, exponential power distribution.

Then, semiparametric models provide greater possibilities for model adjustment wherein response may be one rate; in this case, for cancer mortality, since it allows for non-lineal modelling effects in which variance is not constant; and particularly so, when compared to traditional models that have been employed to date, specifically, the generalized lineal model for Poisson response.


\begin{thebibliography}{3}
\bibitem{Berk} \textsc{Berk, R., \& MacDonald, J.}
\textit{Overdispersion and Poisson Regression}, 2008, Journal of Quantitative Criminology, 24: 269.
\bibitem{Cav} \textsc{Cabanes, A., Vidal, E.,Pérez-Gómez, B., Aragonés, N., López-Abente, G., \& Pollán, M.}
\textit{Age-specific breast, uterine and ovarian cancer mortality trends in Spain: Changes from 1980 to 2006}, 2009, Cancer Epidemiology, 33(169-175).
\bibitem{Cle} \textsc{Clèries, R., Martínez, J., Escribà, J., Esteban, L., Pareja, L., Borràs, J., \& Ribes, J.}
\textit{Monitoring the decreasing trend of testicular cancer mortality in Spain during 2005–2019 through a Bayesian approach}, 2010, Cancer Epidemiology, 34(244-256).
\bibitem{DANE} \textsc{Departamento Administrativo Nacional de Estadística (DANE)}, 
\textit{Demografía y Población}, 2017, Available: http://www.dane.gov.co/index.php/estadisticas-por-tema/demografia-y-poblacion.
\bibitem{Guo} \textsc{Guo, P., \& Li, K.}
\textit{Trends in esophageal cancer mortality in China during 1987–2009: Age, period and birth cohort analyzes}, 2012, Cancer Epidemiology, 36(96-105).
\bibitem{INC} \textsc{Instituto Nacional de Cancerología \& Ministerio de Salud y Protección Social - Departamento Administrativo de Ciencia Tecnología e Innovación en Salud (COLCIENCIAS)}
\textit{Guía de práctica clínica (GPC) para la detección temprana, tratamiento integral, seguimiento y rehabilitación del cáncer de mama}. Sistema General de Seguridad Social en Salud – Colombia Guía Completa - Guía No. GPC-2013-19. 2013.
\bibitem{kv} \textsc{Kv\aa le R., M\o ller, B., Angelsen, A., Dahl, O., Foss\aa , S., Halvorsen, O., Hoem, L., Solberg, A., Wahlqvisti, R. \& Bray, F.}
\textit{Regional trends in prostate cancer incidence, treatment with curative intent and mortality in Norway 1980–2007}, 2010, Cancer Epidemiology, 34(359-367).
\bibitem{Marchant} \textsc{Marchant, D. J.} 
\textit{Epidemiology of breast cancer}. Clinical obstetrics and gynecology, 25(2), 387-392. 1982.
\bibitem{NCI} \textsc{National Center Institute}
\textit{Risk Factors for Cancer}, 2015. Available: https://www.cancer.gov/about-cancer/causes-prevention/risk
\bibitem{Ohshima} \textsc{Ohshima, H. \& Bartsch, H.}
\textit{Chronic infections and inflammatory processes as cancer risk factors: possible role of nitric oxide in carcinogenesis.} Mutation Research/Fundamental and Molecular Mechanisms of Mutagenesis, V(305),\textbf{2}(253-264). 1994.
\bibitem{Pike} \textsc{Pike, M. C., Spicer, D. V., Dahmoush, L., \& Press, M. F.}  
\textit{Estrogens progestogens normal breast cell proliferation and breast cancer risk.} Epidemiologic Reviews, 15(1), 17-35. 1993.
\bibitem{Atlas} \textsc{Piñeros, M., Pardo, C., Gamboa, O. \& Hernández, G.},
\textit{Atlas de mortalidad por Cáncer en Colombia}, Instituto Nacional de Cancerología; IGAC.
3 ed. Bogotá: Imprenta Nacional de Colombia, 2010.
\bibitem{R} \textsc{R Core Team (2016).} 
\textit{A language and environment for statistical computing. R Foundation for
  Statistical Computing, Vienna, Austria. URL https://www.R-project.org/}.
\bibitem{Van} \textsc{Vanegas, L. H.} \& \textsc{Paula, G.A.}
\textit{A semiparametric approach for joint modeling of median and skewness.}. 2014;24: 110–135.
\bibitem{Van_R} \textsc{Vanegas, L. H.} \& \textsc{Paula, G.A.}
\textit{ssym: Fitting semi-parametric log-symmetric regression models. R package version 1.5.7. https://CRAN.R-project.org/package=ssym}, 2016.
\bibitem{WHO} \textsc{WHO},
\textit{International Classification of Diseases. World Health Organization}, 2016. Available: http://www.who.int/classifications/icd/en/
\bibitem{OMS} \textsc{World Health Organization}
\textit{Cáncer}, 2017, Available: http://www.who.int/mediacentre/factsheets/fs297/es/
\end{thebibliography}
\end{document}